# Efficient tenfold up-conversion through steady-state non-thermal-equilibrium excitation

Dafna Granot[1], Nimrod Kruger[1], Assaf Manor[2], and Carmel Rotschild[1,2,3*]
1. Grand Technion Energy Program, Technion – Israel Institute of Technology, Haifa 32000, Israel
2. Russell Berrie Nanotechnology Institute, Technion – Israel Institute of Technology, Haifa 32000, Israel
3. Department of Mechanical Engineering, Technion – Israel Institute of Technology, Haifa 32000, Israel

**Abstract:**
Frequency up-conversion of a few low-energy photons into a single high-energy photon contributes to imaging, light sources, and detection. However, the up-converting of many photons exhibits negligible efficiency. Up-conversion through laser heating is an efficient means to generate energetic photons, yet the spectrally broad thermal-emission and the challenge of operating at high temperatures limit its practicality. Heating specific modes can potentially generate narrow up-converted emission; however, so far such 'hot-carriers' have been observed only in down-conversion processes and as having negligible lifetime, due to fast thermalization. Here we experimentally demonstrate up-conversion by excitation of a steady-state non-thermal-equilibrium population, which induces steady, narrow emission at a practical bulk temperature. Specifically, we used a 10.6μm laser to generate 980nm narrow emission with 4% total efficiency and up-converted radiance that far exceeds the device's possible black-body radiation. This opens the way for the development of new light sources with record efficiencies.

**TOC graphic:**

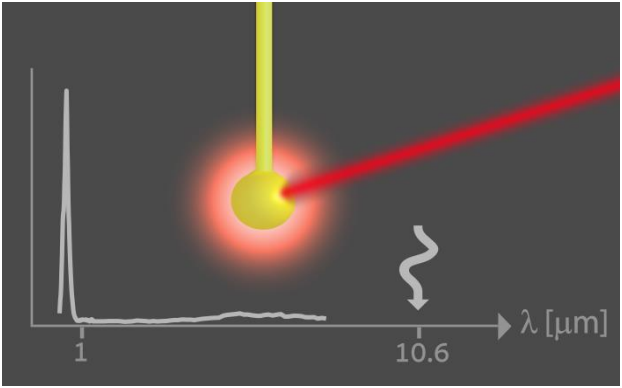

Conventional optical frequency up-conversion methods include coherent (second, third and parametric up-conversion[1–3]) and incoherent (two photon[4,5] and multi-photon absorption[6]) processes which are of major importance in various fields[1,7,8]. Yet, up-conversion of many photons (~10) exhibits negligible efficiency, due to the large momentum mismatch between the pump and the produced frequencies in the parametric process[9], as well as the high intensities required for many photons to interact simultaneously in the multi-photons absorption process. Therefore, the record efficiency of tenfold up-conversion is less than 0.01%, achieved under pulse excitation at intensities of $10^{15}$ W/cm² – many orders of magnitude above currently available *continuous wave* (CW) sources[10].

A naive approach to efficiently generate high optical frequencies is to use a partially coherent low photon energy pump (such as an infrared laser) to increase a body's temperature, thereby generating thermal radiation with an energetic spectral tail in the visible. Although such heating was shown to be highly efficient[11,12], it is challenging to implement, as very high temperatures are needed to produce a considerable amount of energetic photons. Moreover, the resulting broad thermal emission is unsuitable for many applications, and its intensity is limited to the black-body curve. Broad thermal emission can be tailored by various methods[13,14], but successful demonstration of narrow emission at high temperatures has not yet been achieved. An ideal device would operate efficiently at practical temperatures and emit up-converted light at a narrow spectrum. These ideal guidelines can be met if specific modes are heated to a temperature that is much above the bulk temperature. Such non-thermal radiation is characterized by high radiance above thermal emission at specific frequencies. In the case where these "hot-modes" are highly coupled, they equilibrate the excitation between them, and are treated as being in quasi-thermodynamic equilibrium. Their emission is formulated by introducing a scalar chemical potential, $\mu$, into Planck's thermal emission[15–17]:

$$R(\hbar\omega, T, \mu) = \varepsilon(\hbar\omega) \cdot \frac{(\hbar\omega)^2}{4\pi^2\hbar^3 c^2} \frac{1}{e^{\frac{\hbar\omega - \mu}{K_B T}} - 1} \cong$$

$$R_{thermal} \cdot e^{\frac{\mu}{K_B T}}, \qquad (1)$$

Where $R$ is emitted photon flux (photons per second per unit area per solid angle per frequency), $T$ is temperature, $\varepsilon$ is emissivity, $\hbar\omega$ is photon energy, $K_b$ is Boltzmann's constant and $R_{thermal}$ is Planck's thermal emission. Multiplying $R$ by photon energy results in hemispherical radiance, $L = R \cdot \hbar\omega$. Chemical potential, $\mu$, is directly connected to Gibbs free energy: $G = \sum \mu N$ ($N$ being the number of emitted photons with equal $\mu$), and thus to the amount of work that can be done with respect to the bulk temperature. A pump at radiance characterized by a high value of $\mu$ can spontaneously excite modes to emit radiance with lower (yet higher than zero) $\mu$ levels.

Another description for radiance above thermal emission attributes highly excited modes with a brightness-temperature higher than the bulk temperature. Although in such non-equilibrium conditions the term temperature is not well defined, brightness-temperature describes the temperature of a black-body whose emission at a specific frequency equals that of the emitting mode of interest [18–20]. A body under ideal thermal insulation which absorbs radiation at a given brightness-temperature is heated to that temperature. The brightness-temperature $T_b$ for a given bulk temperature $T$, photon energy $\hbar\omega$ and chemical potential $\mu$ is: $T_b = T \cdot \hbar\omega/(\hbar\omega - \mu)$.

Excitation to high brightness-temperature was detected over four decades ago in semiconductors[21], where short pulsed laser generated 'hot-electrons' much above the bandgap, with a brightness-temperature of 3,700K. This excitation was followed by fast thermalization to the bulk temperature[22], which remained at 290K. To the best of our knowledge, the excitation of hot-carriers has been demonstrated only in down-conversion configurations, where the pump photon energy exceeded that of the hot-electrons[10,21]. As far as we know, utilization of such ideas for frequency up-conversion has never been explored, even though thermalization is driven by free-energy (brightness-temperature) and not exclusively by photon energy. That is, a mode with a high brightness-temperature thermalizes to higher frequency modes at lower brightness-temperature as is regularly done in optical refrigeration[23]. Thermalization continues to other modes until the brightness-temperature approaches the bulk temperature at equilibrium.

In a more detailed quantum picture, energy dissipates according to the coupling between modes, which depends, among other parameters, on the density of states as well as on selection rules. An emitter efficiently coupled to the initially excited modes emits spectrally narrow radiation at high brightness-temperature, even if the emitted radiation has a higher frequency than the absorbed frequency.

Experimentally, we used a CW $CO_2$ laser at 10.6μm wavelength to resonantly pump the vibronic states of silica (figure 1a). Their high density of states at wavelengths longer than 8μm induces absorption cross-section of a few microns[24], leading to high brightness-temperatures. Furthermore, Silica has low Density of States (DoS) and therefore barely radiates at wavelengths between 0.4μm to 5μm. We doped the Silica with rare-earth ions, such as Ytterbium ($Yb^{3+}$, emitting at 980nm) or Neodymium ($Nd^{3+}$, emitting at 820nm, 900nm and 1064nm)[11]. These rare-earth emitters are often used in optical refrigeration due to their high quantum efficiency of anti-Stokes florescence, where vibronic excitations are coupled with the radiative transition[23,25]. Using these materials in our experiment allows for efficient cascaded energy transfer from the laser pump to the silica and onwards to the emitter, even though the emission frequency is higher than that of the pump.

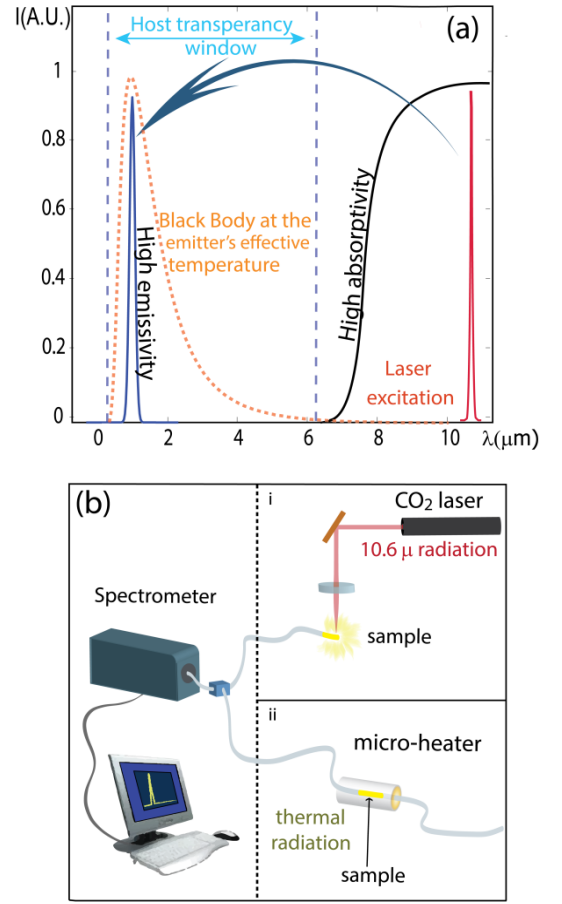

**Figure 1**: Experimental configuration: (a) $CO_2$ laser excited the vibrational modes of the silica (10.6 μm), which were coupled with emitting modes of rare-earth ions at about 1μm. (b) The experimental setup comparing the emission under $CO_2$ excitation to the sample's thermal emission.

Initially, we demonstrated by experiment the non-thermal-equilibrium nature of such excitation, by comparing the emission of the rare-earth doped sample under CW $CO_2$ laser excitation with its thermal emission at high temperature achieved using a furnace (Figure 1b). We connected a 50μm long $Nd^{3+}$ doped silica fiber to a passive silica fiber, and placed it in a micro-heater at 1,400°C, the maximal temperature at which the fiber does not deform. The guided emission was measured at the end of the fiber using a calibrated spectrum analyzer (see the solid red line in figure 2a). The extrapolation of this plot to the expected emission at the melting temperature of silica (1,650°C) is depicted by the purple dotted line in figure 2a (see methods section for details). This served as an upper limit for thermal radiation of the stable device. The solid blue line in figure 2a depicts the emission of an equally measured identical sample under $CO_2$ laser excitation at 0.5W. The radiance under $CO_2$ excitation was measured to be five times higher than the maximal thermal excitation at the melting temperature. We verified that the sample remained stable throughout dozens of experimental runs.

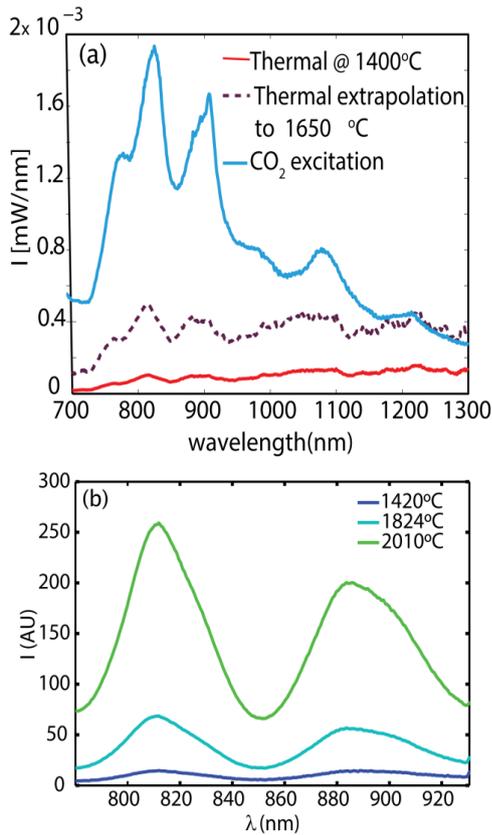

**Figure 2**: Measurements of non-thermal emission: (a) The emission power spectrum around 1μm under CW $CO_2$ excitation (blue line) exceeded that of the measured thermal emission at 1,400°C (red line) and that of the extrapolated thermal emission at the melting temperature of 1,650°C (dotted line). (b) FIR measurements indicated $Nd^{3+}$ maximal brightness-temperature of 2,010°C (green line), much above the maximal possible bulk-temperature.

In addition, we used the *fluorescence intensity ratio* (FIR) method[26] to determine the brightness-temperature of the emitting modes. In this method, the multiple emission peaks of $Nd^{3+}$ (as opposed to $Yb^{3+}$'s single peak) are exploited to ascertain the temperature, by measuring the radiance ratio between two adjacent peaks (see methods section for more details). Figure 2b depicts the device's emission spectrum around 1μm under various $CO_2$ laser intensities below the melting threshold. The green line shows the brightness-temperature at 0.5W excitation that reaches 2,010°C, 360 degrees higher than the bulk melting temperature.

These two different observations indicated that the emitting modes had considerably higher brightness-temperature than the bulk-temperature.

After confirming the non-thermal-equilibrium nature of our up-conversion process we proceeded to examine experimentally its internal and total efficiencies. $Yb_2O_3$-doped $SiO_2$ spherical fiber-tips were produced, with diameters between 30μm and 300μm. At $CO_2$ power level of 684mW, the emission image was captured using a Si CCD camera, which is able to detect wavelengths shorter than 1.1μm, in order to gain information on the size of the emitting area – essential for calculating the radiance. A typical image, with emitting area of $10^{-8}m^2$, is shown in the inset of Figure 3. Next, the sample's spectral radiance between 0.4μm and 11μm wavelength was detected by exciting the sample within an integrating sphere connected to a spectrograph and calibrated against a calibrated lamp and a black-body source at 1,200°C (see methods, section 3). Normalizing the spectral radiance with the emission area resulted in a minimal value for radiance, assuming isotropic emission (see solid lines in figure 3).

Our measurements resulted in a sharp, narrow peak at 980nm attributed to the Ytterbium emission, which included 27% of all the total emittance, indicating relatively high internal efficiency. Residual broad emission, from approximately 5μm to 8μm, originated from silica's highly emissive spectral region and fits the theoretical black-body curve at 1650°C - just below the bulk melting temperature (red dotted line in Fig. 3). The $Yb_2O_3$ emission at 980nm peak was measured as four times higher than this theoretical curve, implying that while the bulk was stable just below 1,650°C, the brightness temperature of the $Yb_2O_3$ was about 2050°C.

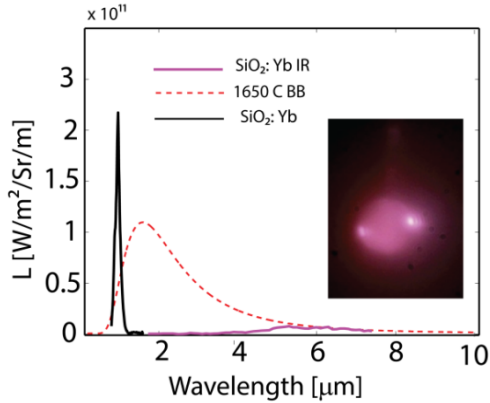

**Figure 3:** Spectral measurements of the $Yb_2O_3$ doped silica under $CO_2$ excitation. Measured power spectrum (black and pink line) indicates 27% internal efficiency and surpasses a theoretical black-body radiance at 1,650°C (dotted red line), the melting temperature of the bulk. This Implies a high brightness-temperature of the $Yb_2O_3$'s emitting modes compared to the bulk-temperature.

Finally, external efficiency was determined by placing the setup in a vacuum chamber at $10^{-3}$ Torr, in order to reduce thermal losses. The $Yb_2O_3$ emission power at the 980nm peak was measured at various CW $CO_2$ power levels. Efficiency was calculated as the ratio between the measured absorbed $CO_2$ power and the measured $Yb_2O_3$ emission power (Fig. 4 dots). The resulting maximal external efficiency approached 4%, where 1.75mW of $Yb_2O_3$ radiation was detected under 49mW $CO_2$ pump intensity.

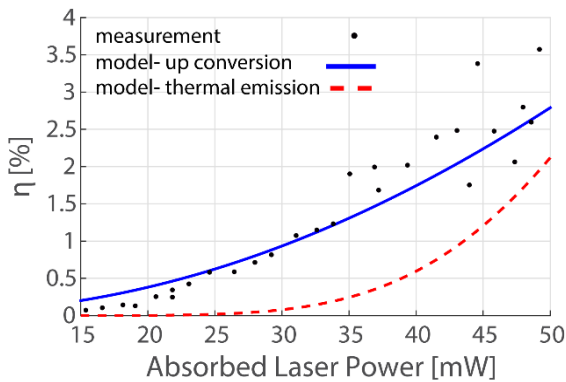

**Figure 4:** Total efficiency of the $Yb_2O_3$ doped silica at various $CO_2$ power levels under vacuum (black dots). The maximal efficiency achieved was over 3.5%, many orders of magnitude better then prior art. The trend fits a simulation (blue line) based on a quantum-thermodynamic model of rare-earths emission. Calculated thermal-radiation (red dashed line) show inferior results.

The efficiency trend is in accord with a model based on Torsello's quantum-thermodynamic theory of thermal pumping and radiative de-excitation in rare-earth oxide materials[27] (Fig.4 blue line), using our experimental parameters [see supporting material]. For these parameters, the model also predicts a maximal possible efficiency of about 4%, slightly above our measured value, where the emission becomes thermal. Moreover, the dashed line in Fig. 4 predicts lower efficiency for an equivalent sample in equilibrium state, confirming the supremacy of the up-conversion over thermal emission.

Further experiment presented in the supporting information section, shows the sample's emission lifetime to be 12-14msec; longer than the typical $Yb^{3+}$ photoluminescence lifetime of about 1msec[28]. This can be explained by the delaying effect of the cascaded energy transfer.

To conclude, we presented a new frequency up-conversion mechanism driven by excitation of hot radiative modes, where the emission under resonant excitation exceeded the maximal possible thermal emission. Specifically, we experimentally demonstrated a tenfold up-conversion from $CO_2$ emission at 10.6μm to 1μm at internal efficiency of 27% and total efficiency of almost 4% – orders of magnitude higher than current state-of-the-art. The efficiency can be further improved using different materials and excitation wavelengths, along with conventional methods for tailoring thermal emission such as photonic bandgap structures. This method may open new directions in various fields of research, such as new light sources and solar energy, where efficiency can be enhanced by up-converting sub-bandgap solar photons to wavelengths where photovoltaics are most efficient.

**Methods**

**Experiment: Up-conversion compared with thermal emission**

**Sample preparation:** a 125μm Nd doped silica optical fiber (CorAcrive) was stripped and connected using an Ericsson fusion splicer to a matched passive silica fiber. The active fiber was then cut to a length of 50μm and connected to another passive fiber, resulting in 50μm active fiber inserted between two long (~50cm) passive fibers.

**Thermal emission measurement**: the fiber was inserted into a high temperature micro-heater (Micropyretics Heater Instruments, Inc), with the Nd doped section in the middle of the oven, held by translation stages at the two outer sides. One side of the fiber was coupled to a monochromator (Andor) with InGaAs camera (Andor), calibrated using a calibration lamp (Oriel Instruments). During the experiment, temperature was gradually elevated to 1,400°C, and the emission at 0.7μm-1.5μm was recorded. The background emission was then measured by pulling the active fiber out of the heater, having only the passive part heated, thus the measured signal originating solely from the heater. Extrapolation of the result to 1,650°C was conducted by multiplying it by $\frac{R(\hbar\omega,1400°C)}{R(\hbar\omega,1650°C)}$, calculated from Eq.1., with µ=0.

**Up-conversion emission measurement**: An identical sample was connected to a passive fiber coupled to the monochromator. The 50µm Nd-doped fiber tip was pumped by CW $CO_2$ laser (Access Lasers) at less than 1W, and measurements were taken at the same conditions as in the previous case. When the $CO_2$ laser was focused on the passive fiber, no emission was detected.

**FIR experiment**
Within a narrow spectral band where the chemical potential is assumed to be equal, the ratio between two adjacent luminescent peaks is defined by Boltzmann distribution:

$$R = C(\Delta E) \cdot e^{-\frac{\Delta E}{K_b T}} \qquad (2)$$

where ΔE is the energy difference between the two levels – 900nm and 820nm – and C(ΔE ) is a constant that can be found by calibration[26].

**Calibration:** In order to ascertain C(ΔE ), an active Nd doped fiber was inserted into a long matching passive fiber as described at section 1, and the active part was put into the micro-heater. One side of the fiber was coupled to the monochromator with the InGaAs camera, while the other side was pumped by a 532nm laser. The temperature was increased gradually, with a measurement taken for each temperature, whereby intensities of 820nm and 900nm were recorded and the ratio between them calculated. A calibration plot was made using data from these measurements ($I_{900}/I_{820}$ Vs T(C)) and C was extracted and found to be 0.4. Without the 532nm pump, the thermal signal was too weak for detection. Eq. 1 and Ref.14 show how photoluminescence has a relative spectrum identical to that of thermal emission.

**Measurements:** Nd doped fiber of the type used for the calibration was pumped by the $CO_2$ laser and its emitted radiation was detected in free space using a Si detector (Ocean Optics). Measurements were taken for different laser powers under 1W. The ratio $I_{900}/I_{820}$ was derived from the measurements and inserted into Eq. 2 to obtain the temperature.

**Efficiency experiments**
**Sample preparation:** Two types of samples were used – the first one was Yb-doped commercial 500µm fibers (CorActive). The second type was a sphere at a fiber tip, made by the following method: 125µm fibers (Thorlabs) were positioned in the focus of a $CO_2$ laser (Synrad). The fiber tip was melted and fed into the focal point until the formation of a sphere. Subsequently, the Sphere was dip-coated in methanol $Yb_2O_3$ nano-crystals in suspension (1ml methanol:100mg $Yb_2O_3$) and then melted again by a short laser expose of about half a second, in order to gain a smooth surface sphere and uniform $Yb_2O_3$ concentration. For smaller spheres, the fibers were etched in an $HF:H_2O$ solution prior to the melting process, until the desired diameter was reached.

**Measurements:** Sample radiance was measured using an integrating sphere (LabSphere for the NIR range and a custom made gold coated sphere for the infrared range). A CW $CO_2$ laser with 2% stability (Access Lasers) at various intensities (measured using a power-meter) was focused onto the sample by a gold parabolic mirror and ZnSe lens. In addition, a calibration lamp (Newport) and a 1,200°C calibrated black-body source (CI-systems) were directed into the sphere for calibration purposes. The luminescence signals were chopped by an optical chopper (Stanford Research Systems) and amplified by a lock-in amplifier (Stanford Research Systems) after passing through a spectrograph equipped with the appropriate gratings for the different spectral ranges (Oriel Instruments). In the near Infrared region, signals were measured by a Ge detector (Judson Technologies) and a InGaAs camera (Andor Technology). In the infrared region between 2µm and 10µm, signals were detected using InSb and MCT (InfraRed Associates) detectors.

**Author information**
Corresponding author-
Email (C.Rotschild): carmelr@tx.technion.ac.il Tel: +927-77-887-1740
Author contributions
The first three authors contributed equally to this paper.
Notes
The authors declare no competing financial interest.

**Acknowledgments**
This report was partially supported by the Russell Berrie Nanotechnology Institute (RBNI) and the Grand Technion Energy Program (GTEP) and is part of The Leona M. and Harry B. Helmsley Charitable Trust reports on Alternative Energy series of the Technion and the Weizmann Institute of Science. We also would like to acknowledge partial support by the Focal Technology Area on Nanophotonics for Detection. A. Manor thanks the Adams Fellowship program for financial support. Prof. C. Rotschild thanks the Marie Curie European reintegration grant for its support.

**Associated content**

Supporting Information Available: lifetime measurements, Semi-empirical model for the up-conversion external efficiency. This material is available free of charge via the Internet at http://pubs.acs.org